\pdfoutput=1

\documentclass[10pt]{article}

\usepackage{listings}    
\usepackage{graphicx}    
\usepackage{subcaption}  
\usepackage{algorithm}   
\usepackage{algorithmic} 
\usepackage{booktabs}    
\usepackage{hyperref}    
\usepackage{xurl}        
\usepackage[htt]{hyphenat} 
\usepackage{microtype} 
\DisableLigatures{}    
\usepackage{adjustbox} 
\usepackage{colortbl}  

\usepackage{caption}
\usepackage{setspace}
\usepackage{multirow}
\usepackage{enumerate}
\usepackage{pdflscape}
\usepackage{pgfplots} 

\usepackage{xcolor}
\definecolor{codegreen}{rgb}{0.25,0.5,0.35}
\definecolor{codegray}{rgb}{0.5,0.5,0.5}
\definecolor{codepurple}{rgb}{0.6,0,0}
\definecolor{backcolour}{rgb}{0.95,0.95,0.92}
\definecolor{colorstring}{rgb}{0.5,0,0.35}
\definecolor{rltred}{rgb}{0.5,0,0}
\definecolor{rltgreen}{rgb}{0,0.5,0}
\definecolor{rltblue}{rgb}{0,0,0.5}
\definecolor{DarkGreen}{rgb}{0.00,0.60,0.00}
\definecolor{ScarletRed}{rgb}{0.80,0.00,0.00}
\definecolor{blizzardblue}{rgb}{0.67, 0.9, 0.93}
\definecolor{green-yellow}{rgb}{0.68, 1.0, 0.18}
\definecolor{dkgreen}{rgb}{0,0.6,0}
\definecolor{gray}{rgb}{0.5,0.5,0.5}
\definecolor{mauve}{rgb}{0.58,0,0.82}
\definecolor{lightgrey}{rgb}{0.90,0.90,0.90}
\definecolor{grey}{gray}{0.75}
\definecolor{light-gray}{gray}{0.80}

\lstdefinestyle{mystyle}{
    escapechar=©, 
	backgroundcolor=\color{backcolour},
    basicstyle=\footnotesize\ttfamily,
   	identifierstyle=\footnotesize\ttfamily,
	commentstyle=\color{codegreen},
	keywordstyle=\color{colorstring}\bfseries,
	numberstyle=\ttfamily\color{codegray},
	stringstyle=\ttfamily\color{DarkGreen},
	breakatwhitespace=false,
	breaklines=true,
	captionpos=b,
	keepspaces=true,
	numbers=left, 
	numbersep=2pt,
	showspaces=false,
	showstringspaces=false,
	showtabs=false,
	tabsize=2
}
\usepackage{xspace}
\newcommand{\evo}{{\sc EvoMaster}\xspace}
\newcommand{\etal}{{\emph{et al.}}\xspace}

\usepackage{boxedminipage}
\newenvironment{result}%
{\smallskip
	\noindent
	\let\emph=\textbf
	\begin{boxedminipage}{\columnwidth}\begin{center}\em}%
		{\end{center}\end{boxedminipage}%
}

\usepackage{ifthen}
\newboolean{showcomments}
\setboolean{showcomments}{true} 

\ifthenelse{\boolean{showcomments}}{
	\newcommand{\nbc}[3]{
		{\colorbox{#3}{\bfseries\sffamily\scriptsize\textcolor{white}{#1}}}
		{\textcolor{#3}{\sf\small$\langle$\textit{#2}$\rangle$}}}
	
}{
	\newcommand{\nbc}[3]{}

}

\usepackage{authblk}                
\usepackage[square,numbers]{natbib} 
\usepackage{amsmath}                
\usepackage{amssymb}                

\usepackage{geometry}
\title{
Fuzzing REST APIs in Industry: Necessary Features and Open Problems

}

\author[1,3]{Andrea Arcuri}
\author[6]{Alexander Poth}
\author[6]{Olsi Rrjolli}
\author[2,1]{Philip Garrett}
\author[2,1]{Juan P. Galeotti}

\affil[1]{Kristiania University of Applied Sciences, Norway}
\affil[2]{University of Buenos Aires and CONICET, Argentina}
\affil[3]{Oslo Metropolitan University, Norway}
\affil[4]{Erciyes University, Türkiye}
\affil[5]{Beihang University, China}
\affil[6]{Volkswagen AG}

\date{}

\begin{document}

\maketitle

\begin{abstract}
REST APIs are widely used in industry, in all different kinds of domains.
An example is Volkswagen AG, a German automobile manufacturer.
Established testing approaches for REST APIs are time consuming, and require expertise from professional test engineers.
Due to its cost and importance, in the scientific literature several approaches have been proposed to automatically test REST APIs.
The open-source, search-based fuzzer \evo is one of such tools proposed in the academic literature.
However, how academic prototypes can be integrated in industry and have real impact to software engineering practice requires more investigation.
In this paper, we report on our experience in using \evo at Volkswagen AG,  as an \evo user from 2023 to 2026.
We share our learnt lessons, and discuss several features needed to be implemented in \evo to make its use in an industrial context successful.
Feedback about value in industrial setups of \evo was given from Volkswagen AG about 4 APIs.
Additionally, a user study was conducted involving 11 testing specialists from 4 different companies.
We further identify several real-world research challenges that still need to be solved.

\end{abstract}


\section{Introduction}

REST APIs are used everywhere, to provide all different kinds of data and functionalities over a network (e.g., internet).\footnote{\url{https://apis.guru}}$^,$\footnote{\url{https://rapidapi.com}}
They are also common when developing backend applications, particularly when using microservice architectures~\cite{newman2021building,rajesh2016spring}.
Nowadays, when interacting with a web page or a mobile app, often one or more REST APIs are involved.
Therefore, the validation and verification of this type of web service is of paramount importance.

Volkswagen AG is a German automobile manufacturer.\footnote{\url{https://www.volkswagen.de}}
As for many enterprises, its IT services rely on REST APIs.
Due to the high cost of thorough testing from professional test engineers, significant effort has been spent to modernize their processes, and leverage what novel techniques and research outputs can provide in this context.
In particular, the use of novel Artificial Intelligence (AI) techniques seems promising.
To enhance the quality of their testing processes and reduce cost, different AI techniques available to the public, like based on LLM (e.g., StarCoder~\cite{li2023starcoder}) and Evolutionary Computation (e.g., \evo~\cite{arcuri2021evomaster}), have already been evaluated at Volkswagen~\cite{poth2025technology}, with some initial success.

In the scientific literature, in the last few years there has been a lot of work on test automation for REST APIs~\cite{golmohammadi2023testing}.
``Fuzz testing''~\cite{zeller2019fuzzing,zhu2022fuzzSuvery,godefroid2020fuzzing} (also known as ``fuzzing'') is a term used to refer to the automated generation of test cases, typically with random or unexpected inputs, to find crashes and security issues in the tested applications.
Several techniques can be used to improve performance (e.g., to cover more parts of  the code of the tested application), e.g., based on AI techniques.
In the literature, several tools (i.e., fuzzers) have been proposed, like the aforementioned \evo.
Most of these tools are open-source, like for example
Restler~\cite{restlerICSE2019}
and
ResTest~\cite{martinLopez2021Restest}.
Any enterprise in the world can download and try out those tools on their REST APIs.

Usually, though, in the scientific literature these tools  have been evaluated only in the ``lab''.
Researchers might design and develop some novel techniques, implement them in a tool, and then carry out experiments on some APIs to evaluate the effectiveness (or lack thereof) of their novel techniques.
Real-world APIs might be used for these experiments, but usually no engineers or QA specialist  in industry would be involved in using and evaluating those tools.
In other words, no ``human aspects'' of introducing fuzzing techniques in industry~\cite{nourry2023human} has been studied so far in literature of testing REST APIs~\cite{golmohammadi2023testing}.

To fill this important gap in the scientific literature, the authors of \evo were eager to start working with the test engineers at Volkswagen AG.
The first interaction happened in October 2023 when the test engineers at Volkswagen contacted the maintainers of \evo with questions about more advanced use cases.

This is what started the ``open exchange''
between the authors of \evo and the test engineers at Volkswagen AG,
in particular with the
Quality innovation Network (QiNET) of the Group IT, which focus on innovations about IT quality management and engineering~\cite{poth2018innovate,poth2023frugal}.
In the literature, there are many types of academia-industry collaborations~\cite{garousi2019characterizing,garousi2016challenges,garousi2017industry,garousi2017}.
Bridging the gap between academia and industry is an important research endeavour, that can provide benefits for both parties.
As such, this is explicitly mentioned in the documentation of \evo regarding how people can contribute.\footnote{\url{https://github.com/WebFuzzing/EvoMaster/blob/master/docs/contribute.md}}

In this paper, we report on the first two years of this exchange between the developer team of \evo and QiNET at Volkswagen AG.
The Volkswagen engineers evaluated \evo in an industrial setup which opened additional usage scenarios for the \evo team.
We discuss all the technical details and features that have been implemented to be able to better integrate an academic fuzzer such as \evo into the testing practices of a large enterprise such as Volkswagen AG~\cite{poth2022integration}.
We discuss all the challenges that we have overcome, and highlight several challenges that still need addressing.
Many of these challenges are not specific to \evo, but would likely be relevant to all other kinds of fuzzers and test generators whose authors want to have an impact on industrial practice.
Empirical studies have been carried out to show and quantify the actual benefits that a tool such as \evo can provide to practitioners in industry.
This included 4 APIs at Volkswagen, and the different test engineers responsible for them, besides a user-study involving 11 AI-test specialists from 4 different companies.

This paper is an extension of a conference version~\cite{icst2025vw}.
In that work, the first year of the exchange was discussed.
To make clear the distinction with what presented in this paper and what already presented in~\cite{icst2025vw}, we explicitly divide subsections between these two years.
Note that some challenges identified in~\cite{icst2025vw} have now been solved, and presented so here as contribution of the second year.
The empirical study has also been extended with more APIs and with a new user-study involving 11 AI-test specialists from 4 different enterprises outside of Volkswagen AG.
Furthermore, we added a discussion on other important features requested by the community but that were not relevant or of major interest for the  Volkswagen AG Group IT Test \& Quality Assurance (TQA), which is focused on black-box testing.

Also note that our previous work in~\cite{poth2025technology} evaluated at Volkswagen AG different openly available test generators, including \evo, \emph{``as they are''}, \emph{before} an exchange was started in 2023.
In other words, this work can be considered a follow up of~\cite{poth2025technology}, discussing what has been identified and solved during two years of this academia-industry exchange.

In particular, in this industry-report we aim to shed light on these important research questions:

\begin{enumerate}

\item[]{\bf RQ1}: Why choosing \evo for an academia-industry evaluation and exchange instead of other tools?

\item[]{\bf RQ2}: What features were needed to integrate a fuzzer like \evo in industrial testing processes such as for example at Volkswagen AG?

\item[]{\bf RQ3}: What is missing from the generated tests of \evo compared to the existing manually written test suites?

\item[]{\bf RQ4}: What are the current major challenges and most needed features?


\item[]{\bf RQ5}: Which important REST API fuzzing aspects were not relevant at Volkswagen setting?

\end{enumerate}

The paper is organized as follows.
Section~\ref{sec:relatedwork} starts from discussing related work.
Section~\ref{sec:rq1} answers RQ1, followed by
RQ2 in Section~\ref{sec:rq2},
RQ3 in Section~\ref{sec:rq3},
RQ4 in Section~\ref{sec:rq4} and
RQ5 in Section~\ref{sec:rq5}.
Threats to validity are discussed in Section~\ref{sec:threats}.
Finally, Sections~\ref{sec:conclusions} concludes the paper.

%

\section{Related Work}
\label{sec:relatedwork}

In the literature of software testing, there have been several studies involving the evaluation of academic techniques in industry.
These included topics such as
unit testing~\cite{seip2017industrial},
testing of embedded systems~\cite{zhang2018smartunit},
user interface testing
for ERP applications~\cite{brunetto2021introducing}
and for mobile applications~\cite{sapienz2018,wang2018empirical},
mutation testing~\cite{beller2021would,petrovic2018state},
fault debugging~\cite{zhou2018fault}
and flaky tests~\cite{Godefroid2019Flaky}.
This work extends such current body of knowledge by addressing the important topic of fuzzing REST APIs in industry.

In the last few years, due to its importance in industry, lot of research has been carried out on automated testing of REST APIs~\cite{golmohammadi2023testing}.
Several tools have been presented in the literature, such as for example (in alphabetic order):
ARAT-RL\cite{kim2023adaptive},
AutoRestTest~\cite{kim2025autoresttest}
bBOXRT~\cite{laranjeiro2021black},
DeepREST~\cite{corradini2024deeprest},
EmRest~\cite{xu2025effective},
\evo~\cite{arcuri2025tool},
LLamaRestTest~\cite{kim2025llamaresttest},
Morest~\cite{liu2022icse},
ResTest~\cite{martinLopez2021Restest},
RestCT~\cite{wu2022icse},
Restler~\cite{restlerICSE2019},
RestTestGen~\cite{viglianisi2020resttestgen}
and
Schemathesis~\cite{hatfield2022deriving}.

However, industry reports and empirical studies with human subjects are a rarity in the literature of REST API testing~\cite{golmohammadi2023testing}.
This paper fills this important gap in the research literature.
Besides our previous work at Volkswagen AG~\cite{poth2025technology,icst2025vw}, we are aware of only two other cases.
To study its use for fuzzing RPC APIs in industry, the authors of \evo did a series of user studies in industry with large enterprises such as Meituan~\cite{zhang2025fuzzing}.
In the first of such user studies, RPC APIs were run behind REST APIs, as RPC fuzzing was not supported yet at that time in \evo.
In~\cite{sartaj2025rest}, Sartaj \etal studied the application of five different REST fuzzers on 17 APIs regarding Healthcare Internet of Things, in collaboration with the public administration of the Oslo Kommune Helseetaten.

\evo is a search-based fuzzer, open-source on GitHub since 2016.\footnote{\url{https://github.com/WebFuzzing/EvoMaster}}
It is a mature tool~\cite{arcuri2018evomaster,arcuri2021evomaster,arcuri2023building,arcuri2025tool}, originally designed for search-based white-box testing of REST APIs~\cite{arcuri2019restful}.
However, it has been extended to support black-box testing~\cite{arcuri2020blackbox}, as well as other types of web services such as GraphQL~\cite{belhadi2023random} and RPC~\cite{zhang2023rpc}.
For white-box testing, it uses several advanced search-based techniques,
such as hyper-mutation~\cite{zhang2021adaptive},
testability transformations~\cite{arcuri2021enhancing,arcuri2024advanced},
and support for effectively handling the APIs' environment such as SQL databases~\cite{arcuri2020sql}, MongoDB databases~\cite{ghianni2026search} and external web services~\cite{seran2025handling}.
Besides being able to generate tests cases to detect crashes (e.g., HTTP 500 status code), \evo can also be used to validate security properties, like detecting faults related to access control policies~\cite{arcuri2025fuzzing}.
\evo has been among the best performing tools in empirical comparisons~\cite{Kim2022Rest,zhang2023open,sartaj2025rest,sahin_2025_wfc,arcuri2026sbft}.

\section{RQ1: First Steps}
\label{sec:rq1}

In 2022, at the IT Test \& Quality Assurance (TQA)
 a process was started to check and evaluate possible Artificial Intelligence (AI) techniques to enhance their testing activities in the Group IT.
Discovering \evo was part of such activities.
The tool came out when querying internet search engines (e.g., Google), using terms such as ``AI'' and ``testing''.
Few tools were found this way.
The choice of trying out \evo was based on some specific properties:

\begin{itemize}

\item the tool or component should be fast to integrate and simple to scale in a cloud environment, like for example open-source solutions do.
        \evo is open-source since its first commit in 2016, with LGPL licence.
      However, enterprises might prefer more permissive licenses, such as MIT and Apache, especially if they want to make modifications and extensions to such tools for internal use.
      The choice of an open-source license for a research prototype is not trivial, as discussed for example in~\cite{arcuri2023building}.

\item the tool should be ``\emph{AI-hype compatible}''. In other words, it should explicitly states it uses AI techniques.
    White-box \evo is based on Evolutionary Computation, as explicitly stated in its documentation.

\item the tool should be actively maintained, ideally with at least three active developers,
    and should had been around for some years. This provides more trust for its maintenance in the future.
    As of 2022, \evo was already available open-source for 6 years, with several active code contributors. As of 2026, there are at least 15 code contributors with 50 or more Git commits to \evo.\footnote{\url{https://github.com/WebFuzzing/EvoMaster/graphs/contributors}}


\item the outputs of the tool should be easily integrated in the current testing environments. In the case of Volkswagen AG, this means being able to run the generated test suites with Java and JMeter. As \evo can output tests in JUnit format, this was not a problem.

\item the tool should have given good benchmark results. Engineers at Volkswagen AG checked some academic papers to see how different tools fare with each others.

\end{itemize}

\evo was the  tool that matched all these criteria best.
There are some other tools that ``look interesting'' for the engineers at Volkswagen, such as
Restler~\cite{restlerICSE2019},
Schemathesis~\cite{hatfield2022deriving}
and
CATS.\footnote{\url{https://github.com/Endava/cats}}
However, they did not have the ``AI'' label.
As such, \evo was preferred.

\begin{result}
{\bf RQ1}: the maturity and health of a product or project are key elements to decide whether to invest time to try it out.
To increase adoption in industry, it is recommended to explicitly provide such information in the documentation of the tools.
Specifying which techniques a tool employs can be a major deciding factor, especially when ``hype'' and industry trends are involved (e.g., related to AI).
\end{result}

\section{RQ2: Implemented Features}
\label{sec:rq2}

\begin{table}[!t]
   \centering
    \small
    \caption{
    Summary of the main features to enable a better use of \evo in industrial enterprise settings.
    They are listed in the order in which they will be discussed in the text.
    }
    \label{tab:features}
        \begin{tabular}{ll}\\
        \toprule
        Name &  Benefit(s) \\
        \midrule
        Domain Expert Inputs: Examples &  Ensures that specific business aspects are validated \\
        Domain Expert Inputs: Links    &  Ensures that specific business aspects are validated \\
        Schema Validation              &  Reduces human mistakes \\
        Authentication                 &  Reduces manual effort \\
        Computational Resources        &  Optimizes available resources \\
        Coverage Criteria              &  Produces better test suites \\
        Databases                      &  Produces better test suites \\
        Rate Limiter                   &  Optimizes available resources \\
        \midrule
        Example Objects                & Ensures that specific business aspects are validated \\
        Multi-File Schemas             & Supports multi-stakeholder organizations for large APIs \\
        Test Summaries as Comments     & Makes test validation more efficient \\
        Test Naming                    & Makes test validation more efficient \\
        DTO to Support Test Modifications & Makes test validation more efficient \\
        Automated Data Cleanup         & Supports re-run of tests \\
        \bottomrule
        \end{tabular}
\end{table}

In this section, we discuss all the major features needed by the test engineers at Volkswagen AG to enhance the usability and effectiveness of \evo on the testing of their APIs.
These are summarized in Table~\ref{tab:features}.

\subsection{Features Implemented in the First Year}

\subsubsection{Domain Expert Inputs: Examples}
Fuzzers have limitations, especially black-box ones.
The most important needed feature was a way for the test engineers to provide help to the fuzzer to obtain better results.
Ideally, a fuzzer should be fully automated.
But, as their outputs are still not as good as manually developed test suites~\cite{poth2025technology}, any help that can be exploited could be useful.
But, this is only as long as it does not take too much time to provide and prepare such input data and verify the outcomes.
For example, this could be something as simple as providing valid ids of existing users in the database.

For each parameter of each endpoint, there was a need to provide a ``dictionary'' of meaningful data, based on the actual content of the database in the test environment at Volkswagen.
From a technical perspective, this is nothing too complex.
However, one issue here was to decide on which way and format such test data should be provided as input to \evo.
Creating a custom format, e.g., a custom configuration file with the test data, was a possibility.
However, it is not ideal for two main reasons:
(1) out of the box, IDEs would not support such a new format (e.g., for auto-complete features and validation);
(2) a custom format would not be supported by other fuzzers, and so any effort spent there would have to be repeated when using a different fuzzer.
If there was no other option, a custom format would be a necessity.
However, one alternative is to provide such test data directly in the OpenAPI schemas.
The OpenAPI schemas provide fields called \texttt{example} and \texttt{examples} to specify possible examples of how parameters could look like.
These entries can be used to provide the needed test data.

At that time, \evo did not have any heuristic to exploit \texttt{example} data specified in the OpenAPI schemas.
This is implemented now since version 3.0.0~\cite{zenodo300evomaster}.
Given a certain probability, input is not generated at random, but taken from the provided dictionary (i.e., the values declared in the \texttt{example} entries), if any is specified.
A further benefit here is that any other tool that can exploit \texttt{examples} entries would be able to directly  use such test data.
For example, another fuzzer that explicitly states that it supports \texttt{example} entries is RestTestGen~\cite{corradiniautomated2022}.

\subsubsection{Domain Expert Inputs: Links}
A further issue in testing REST APIs is to identify dependencies among endpoint operations.
For example, the needed input for an endpoint $Y$ might require to come from the output of an endpoint $X$.
In the literature, to improve performance several techniques have been designed to try to infer possible dependencies among operations, like the Operation Dependency Graph used in RestTestGen~\cite{corradiniautomated2022},
or our own heuristics based on template test actions~\cite{zhang2021resource}.

\begin{figure}
\begin{lstlisting}
paths:
  "/api/links/create":
    post:
      tags:
      - bb-links-application
      operationId: postCreate
      responses:
        '200':
          description: OK
          content:
            "*/*":
              schema:
                "$ref": "#/components/schemas/BBLinksDto"
          links:
            LinkToGetUser:
              operationId: getUser
              parameters:
                path.name: "$response.body#/data/id"
                query.name: BAR
                code: "$response.body#/data/code"
  "/api/links/users/{name}/{code}":
    get:
      tags:
      - bb-links-application
      operationId: getUser
      parameters:
      - name: name
        in: path
        required: true
        schema:
          type: string
      - name: name
        in: query
        required: false
        schema:
          type: string
      - name: code
        in: path
        required: true
        schema:
          type: integer
          format: int32
\end{lstlisting}
\caption{\label{fig:schema}
Extract from an OpenAPI schema of an artificial API example, with a \texttt{links} definition.
}
\end{figure}

As for any heuristics, there is no guarantee that they would be helpful on all different types of schemas and styles in which API are designed.
However, as of version 3.0, OpenAPI schemas can now support the definition of ``links''.\footnote{\url{https://swagger.io/docs/specification/links/}}
Links allows to specify that any kind of output from an endpoint could be used as a meaningful input to another endpoint.
Figure~\ref{fig:schema}
shows a simple, artificial example of a link definition.
This example is one of the end-to-end tests used in \evo to verify its functionalities~\cite{arcuri2023building} (in particular, the end-to-end test called \texttt{BBLinksEMTest}).
In a link definition, several values could be defined, which can be constant or be any field or value from the request/response of the linked endpoints.

\begin{figure}
\begin{lstlisting}[language=java]
@Test(timeout = 60000)
public void test_2() throws Exception {

   ValidatableResponse res_0 = given().accept("*/*")
         .post(baseUrlOfSut + "/api/links/create")
         .then()
         .statusCode(200);
   String link_0__data_id = res_0.extract().body().path("data.id").toString();
   String link_0__data_code = res_0.extract().body().path("data.code").toString();

   given().accept("*/*")
        .get(baseUrlOfSut + "/api/links/users/" + link_0__data_id + "/" + link_0__data_code + "?name=BAR")
        .then()
        .statusCode(200);
}
\end{lstlisting}
\caption{\label{fig:test}
Example of generated test showing the dynamic use of a link, based on the OpenAPI schema defined in Figure~\ref{fig:schema}.
}
\end{figure}

If a fuzzer can exploit such link definitions, then test engineers have incentive to spend time adding those links to the schemas of the APIs they are testing.
Furthermore, regardless of testing concerns, adding links improve the quality of the schema and therefore as well the quality of the documentation of the API.
However, supporting OpenAPI links has been far from trivial.
 Not only there are several ways to define links, but also more importantly the dynamic nature of link handling must be maintained in the generated tests.
 Otherwise, it would not be useful, as likely ending up producing flaky tests.
 Figure~\ref{fig:test}
 shows an example of generated test on an API based on the schema defined in Figure~\ref{fig:schema}.
 As we can see in that generated test, two values are dynamically extracted from the first HTTP request and used as input in the second request.

We do not know if any other fuzzers support OpenAPI links.
Still, what is interesting here is that the use of fuzz testing can impact how schemas are extended and updated by practitioners.

\subsubsection{Schema Validation}
This feature was strongly related to the handling of \texttt{examples} and \texttt{links} entries in OpenAPI schemas.
When we implemented the support for \texttt{examples} and \texttt{links} in \evo,
the first feedback from an industrial perspective showed that it was not enough,
as those entries were not used during the search.
First suspect was a fault in these new functionalities, which would not be unexpected.
However, it turned out that the used schemas were invalid.

\begin{figure}
\begin{lstlisting}
paths:
  "/api/links/create":
    post:
      tags:
      - bb-links-application
      operationId: postCreate
      responses:
        '200':
          description: OK
          content:
            "*/*":
              schema:
                "$ref": "#/components/schemas/BBLinksDto"
        links:  ©\label{line:links}©
            LinkToGetUser:
              operationId: getUser
              parameters:
                path.name: "$response.body#/data/id"
                query.name: BAR
                code: "$response.body#/data/code"
\end{lstlisting}
\caption{\label{fig:faulty}
Faulty definition of endpoint, based on schema from Figure~\ref{fig:schema}.
}
\end{figure}

To better explain this issue,
consider Figure~\ref{fig:faulty}
that shows a faulty variant of the OpenAPI schema defined in Figure~\ref{fig:schema}.
Can the reader spot the problem?
That schema is a syntactically valid YAML file, but not an extract of a valid OpenAPI schema.
Here, the entry \texttt{links} at Line~\ref{line:links}
is invalid, because defined under the object \texttt{responses} and not under \texttt{`200'}.
Without any check or warning, such \texttt{links} definition would be ignored.

If a schema is written manually with just the support of a text editor or an IDE, then this type of issues can be easily missed.
Editors that have custom support (e.g., via plugins) for OpenAPI are necessary to avoid this kind of problems.
Still, as we found out, even a valid OpenAPI schema can be problematic for a fuzzer.
For example, the type of the entry \texttt{example} in a Schema Object definition is \texttt{Any}.\footnote{\url{https://swagger.io/specification/\#schema-object}}
This means that inserting wrong types for an example (e.g., not matching the type of field the example is for, like an array of strings for an integer value) could be silently ignored by an editor/IDE, even if it has support for OpenAPI validation.

To handle this issue,  \evo was extended to validate the input OpenAPI schemas.
If there is any issue, we make sure to print meaningful warning messages to the users, inviting them to fix those issues to enable a more performant fuzzing experience.
Even if a schema has issue, \evo would not crash, but rather it will try its ``best-effort'' to fuzz the target API, exploiting all the information it can have access to.

\subsubsection{Authentication}
In most  cases, real-world APIs require some sort of authentication to identify the user that is making the requests.
To be able to effectively fuzz an API, authentication information for some existing users must be provided.
There can be several different types of authentication mechanisms, typically based on some sort of userid and password provided by the user.

Among the different types of authentication mechanisms, the two most commons that we have experienced in practice are:

\begin{enumerate}

\item[] {\bf Static}. A secret is sent at each HTTP request, either in a HTTP header, query or path parameter.
        The secret could be the combination of userid/password, or a random string uniquely associated to the user.

\item[] {\bf Dynamic}. A secret, typically userid and password, is sent to a ``login'' endpoint. If the secret is valid,
    then such endpoint will return a body payload with an authentication ``token''.
    Such ``token'' can then be used as a secret like in the ``Static'' case, i.e., can be sent in a header, query or path parameter.
    The difference here is that the ``token'' might be short-lived, i.e., only valid for some minutes or hours.

\end{enumerate}

Supporting authentication in fuzzing is not a trivial task.
That can be a reason why few fuzzers seem to have only limited or no support at all for it~\cite{zhang2023open,sahin_2025_wfc}.

Black-box \evo already supported the ``Static'' case, with for example options such as \texttt{--header} to specify static HTTP headers sent on each request.
However, handling the ``Dynamic'' case was not supported for black-box testing.
At the beginning, engineers at Volkswagen had to make ``manual'' calls to the login endpoints, and then use the result with \texttt{--header} option when starting \evo.
This could be very frustrating, especially in cases in which API tokens only had a five minute lifespan.

Supporting the ``Dynamic'' case is more complicated.
There is the need to specify several options, like which endpoint to call, how to extract the auth information, and how then to use it in all the following requests.
And this has to be done dynamically, e.g., the generated JUnit test suite files will still then need to be able to do the same.
Once all this information is provided, then it is up to the tool (i.e., \evo in this case), to use such information to collect the needed authentication tokens, and use them during the fuzzing.

This latter part was already implemented in \evo, for the support of white-box testing.
There, authentication information can be specified in the so called \evo \emph{driver} classes~\cite{arcuri2018evomaster}.
Those are classes written in Java or Kotlin, where the user can specify several options, especially how to start, stop and reset the API.
Among those options, there is also the possibility to specify authentication information.
Several utilities are provided to simplify such task.

Doing the same for black-box testing is not straightforward, as there is no Java/Kotlin driver class.
Providing all needed information on the commandline (e.g., with parameters such as \texttt{--header}) did not look like a good idea, as there would be many parameters to set.
In the end, from a usability perspective, the choice was made to define \emph{configuration files} to specify authentication information.
These files can be written in either YAML or TOML format.
We explicitly decided not to support the JSON format, as that format is in our opinion not sufficient for configuration files, as it does not support writing comments in it.

\begin{figure}
\begin{lstlisting}
[[auth]]
name="logintoken"
[auth.loginEndpointAuth]
endpoint="/api/logintoken/login"
payloadRaw= """
{"userId": "foo", "password":"123"}
"""
verb="POST"
contentType="application/json"
[auth.loginEndpointAuth.token]
headerPrefix="Bearer "
extractFromField = "/token/authToken"
httpHeaderName="Authorization"
\end{lstlisting}
\caption{\label{fig:auth}
Example of TOML configuration file for authentication using tokens extracted from a \texttt{login} endpoint, and that can then be
 sent as \texttt{Bearer} in the HTTP \texttt{Authorization} header.
}
\end{figure}

\begin{figure}
\begin{lstlisting}[language=java]
@Test(timeout = 60000)
public void test_0() throws Exception {

   final String token_logintoken = "Bearer " +
      given()
      .contentType("application/json")
      .body(" { " +
                  " \"userId\": \"foo\", " +
                  " \"password\": \"123\" " +
                  " } ")
      .post(baseUrlOfSut + "/api/logintoken/login")
      .then().extract().response()
             .path("token.authToken");

  given().accept("*/*")
      .header("Authorization", token_logintoken)
      .get(baseUrlOfSut + "/api/logintoken/check")
      .then()
      .statusCode(200)
      .assertThat()
      .contentType("text/plain")
      .body(containsString("OK"));
}
\end{lstlisting}
\caption{\label{fig:token}
Example of generated test showing how authentication tokens can be dynamically retrieved and used in following HTTP calls.
}
\end{figure}

Once written, these authentication configuration files can then be given as input to \evo when it starts the fuzzing.
Figure~\ref{fig:auth} shows an example of such a configuration file on an artificial API,
used as end-to-end test in \evo~\cite{arcuri2023building} (in particular, the test \texttt{BBAuthTokenEMTest}).
Figure~\ref{fig:token}
shows a generated test in which such info is used to authenticate a user.

The importance of having configuration files for authentication information goes beyond \evo.
It is a major issue when comparing fuzzers in empirical studies, as each one would have its own custom system (if any).
Furthermore, for practitioners in industry, it makes harder to try out different fuzzers on real-world APIs, as each fuzzer would need to be configured separately.
For these reasons, since our work in~\cite{icst2025vw} to support the authentication needs at Volkswagen AG,  we have extracted this authentication specification in its own separated ``standard'' called WFC (Web Fuzzing Commons), independently of \evo~\cite{sahin_2025_wfc}.
The goal was to enable other fuzzers to support the same kind of authentication configurations.

\subsubsection{Computational Resources}
In a large organization, there can be many, many APIs that need testing.
And, even for the same API, there can be several different versions that might need testing.
Then, this testing process has to be repeated for each new release, for each API.
All this requires a non-trivial amount of computational resources.
When dealing with testing at scale, how to best to use these computational resources becomes critical.

One of the first questions when using a fuzzer is for how long to run it.
Should it be for just 30 seconds? 1 hour? or 48 hours?
We have no answer for this.
A user would want results as soon as possible.
But, a fuzzer would still need some time to be able to produce good results.
Also, the time required to properly fuzz an API would likely depend on its size and complexity.
In industry,
there is the need for a sustainability design perspective~\cite{poth2024sustainable} to make all components, including a fuzzer such as \evo, in the scaled setup ``ready''.

For white-box testing \evo, we usually recommend to run it at least 1 hour.
This can be set with the parameter \texttt{--maxTime}.
However, the value ``1-hour''  is just an educated guess.
For black-box testing, it might be more trickier to give a time recommendation.
Of course, the longer the better, but it is likely that the fuzzing process could stagnate soon if there is no white-box code heuristic that can be exploited to generate better tests.
In other words, one might not see much improvements after 5 or 10 minutes (this of course all depends on the tested API).

To take this feedback and insight into account, we introduced a new parameter \texttt{--prematureStop}, where a timeout can be  specified.
If there is no improvement (i.e., no new testing target is covered) within the timeout, then the fuzzing process is prematurely stopped.
For example, using something like \texttt{--maxTime 1h} and \texttt{--prematureStop 10m} would mean that the fuzzing will run for one hour, but, if no improvement is reached at any point in time within the last 10 minutes, then the fuzzing process is stopped.
Doing something like this enables testers to allow \evo to run for its needed time, but, if for any reason it looks like it gets stuck and does not provide any further improvement, then \evo can be automatically stopped.
Stopping \evo prematurely enables releasing computational resources that could be used to fuzz other APIs.

Another needed feature related to the handling of computational resources was the ability to fuzz only subsets of an API.
In an OpenAPI schema there can be many declared endpoints, sometimes in the order of hundreds.
There might be the need to test only  some of those endpoints (e.g., related to some newly introduced functionalities), or having \evo run in parallel on several different processes/servers, each one testing in parallel a subset of the API.
Supporting such industrial needs was relatively easy, as it was not particularly difficult to provide filtering options to specify which endpoints to test.
This can  for example be based on the prefix of the endpoint paths (e.g., only use endpoints that start with a \texttt{/v3/items*}, by using the option \texttt{--endpointPrefix}), or by ``tags'' specified in the OpenAPI schema (e.g., using the option \texttt{--endpointTagFilter} to filter all endpoints that have the specified tags declared in their schema).

\subsubsection{Coverage Criteria}
Since its inception in 2016, \evo has been first and foremost a \emph{white-box} search-based fuzzer for APIs running on the JVM.
The support for black-box testing was introduced years later, in 2020~\cite{arcuri2020blackbox}.
Still, the black-box mode was introduced only for demonstration purposes, as it is much easier to use (e.g., no need to first write any driver class).
However, currently Volkswagen is more interested in black-box testing, so it was the right time to improve \evo's black-box capabilities in the context of system testing.

The first ``shocking'' revelation was how few tests were generated by \evo.
During a fuzzing session of one hour, depending on the API, hundreds of thousands of HTTP calls can be made by a fuzzer.
For example, a 5ms per call would result in 720 000 calls in one hour.
However, no sane fuzzer would output test suites with hundreds of thousands of HTTP calls.
What is outputted at the end of the fuzzing process is supposed to be a minimized set, including all the most relevant tests.
However, ``relevance'' here all depends on what is defined as coverage criteria.

Besides fault finding, the coverage criteria for white-box \evo are based on code metrics.
For example, on line and branch coverage.
If any new test evolved during the search triggers the execution of a new line in the source code of the tested API, then we make sure such test will be part of the final output test suite.
Unfortunately, this is not possible for black-box testing, as no code metric is collected (if it was, then it would not be black-box).
Black-box coverage criteria were based only on combinations of endpoints and returned HTTP status code.
In other words, we save each test that can return a status code not seen before on each specific endpoint in the API.
Unfortunately, this turned out to be quite limited.

To address such a major limitation, \evo was extended with the set of black-box coverage criteria defined in~\cite{martin2019test}.
This includes for example considering all boolean combinations of optional query parameters (i.e., on and off), as well as covering all values of input enums.
However, we made two major extensions to those coverage criteria defined in~\cite{martin2019test}:
(1) in the coverage targets, we included all \texttt{example(s)} and followed \texttt{links} declared in the schema, as those are the input data the testers are interested into and want to see in the generated tests;
(2) for each coverage target per endpoint, we check when it is covered regardless of the returned status code, and as well as when the returned code is in the 2xx success family.

This latter point requires some explanation.
Assume two query parameters for an endpoint $Z$: the string $X$ and the enumeration $Y$ with 10 possible values.
If $X$ has some constraints that are not satisfied (e.g., a regex), then the API returns 400 when calling $Z$ with such invalid input, regardless of the value of $Y$.
Having in the final test suite 10 test cases for each different value of $Y$ when the endpoint  $Z$ anyway returns a 400 due to invalid $X$ would be of little interest, as those values of $Y$ have no impact on the test.
On the other hand, if the endpoint returns a 2xx, then such values of $Y$ might cover different execution paths inside the API.
Still, it might be possible that an endpoint $Z$ is never covered with a 2xx during the search.
In those unlucky cases, it might still be worthy to have different values of $Y$ in the generated tests.

\subsubsection{Databases}
Databases play a critical role in system testing.
The data contained in them influence what can be tested from the API endpoints.
These test environments are often production comparable systems with a dedicated test-data set in their databases, to be able to run and stimulate the tests of the system.
To the best of our knowledge, it is a common practice in industry to test APIs in a test environment with such anonymized or synthetic data.
This is the case for example for the testing in the context of TQA, and in all the other companies the \evo team has worked in and collaborated with.
Using databases that are copies of production data snapshots is usually not possible, due to privacy and legal issues (e.g., GDPR in Europe).

The data in these databases are a potential source of useful inputs that can be exploited.
For example, consider testing a \texttt{GET} operation on the path \texttt{/products/\{id\}}.
If there is no operation to create a new product (e.g., this is done by a separated admin API), without knowing any valid \texttt{id} entry, it would be difficult to test any scenario besides the trivial 404 (i.e., resource not found).
Having testers adding such values as \texttt{examples} entries is possible, but ideally such manual interventions should be minimized, if possible.

For white-box testing, this is not a problem, as a scenario like this can be trivially solved with SQL handling techniques as introduced in~\cite{arcuri2020sql}.
White-box \evo can analyze all interactions with SQL databases~\cite{arcuri2020sql} (as well as MongoDB~\cite{ghianni2025search}), and then use all different kinds of taint analyses techniques~\cite{arcuri2021enhancing,arcuri2024advanced} to generate input data that make sure each SQL SELECT command done in the API returns data.

To try to exploit existing database information in black-box testing as well, we implemented the same kind of \emph{Response dictionary} technique introduced in~\cite{corradiniautomated2022}.
In simple terms, at a high level \texttt{GET} operations on collections (e.g., \texttt{GET:/products}) can provide info on what present in the database.
From those returned payloads, the ids of the resources can be inferred (with different types of string-matching algorithms and heuristics~\cite{corradiniautomated2022}).
Those collected ids at runtime can then be stored in a dictionary, and used afterwards for the endpoints that require ids as input (e.g., \texttt{PUT:/products/\{id\}}).

This approach might work well when testing remote APIs on internet, where login is done with a newly created user account made only for academic experiments.
However, when dealing with test environments shared by possibly several testers, and having existing manual test cases relying on some specific data, having a fuzzer starting to delete entry by doing \texttt{DELETE:/products/\{id\}} with admin credentials is unwise.
A test environment can still be reset and rolled-back.
But, doing that at each fuzzing execution might not be a viable strategy.
To address this issue, in our implementation the inferred ``Response dictionary'' is  exclusively used only for read operations (i.e., \texttt{GET}).
This is an important problem, which we will go back to in Section~\ref{sec:rq4}.

\subsubsection{Rate Limiter}
When fuzzing an API, the more test cases can be evaluated, the better results can be expected at the end.
However, sending as many requests as possible toward an API could put major strain/stress on its operations.

When academics perform empirical experiments on open-source APIs locally on their machines (e.g., from the WFD corpus~\cite{sahin_2025_wfc}), this is not a problem.
However, when testing an API on the internet (e.g., as done in~\cite{corradiniautomated2022}), this can become problematic.
For example, to prevent and mitigate Denial-of-Service (DoS) attacks, the API could stop replying to requests from those fuzzers.
A solution here is to have a ``rate limiter'', i.e., a way to specify how many requests to send at most given a specified amount of time.
For example, in \evo there is the option \texttt{--ratePerMinute}.
A value like $N=30$ would mean that, within a minute, no more than 30 HTTP requests are sent, i.e., at most 1 request every 2 seconds.
When a request takes $X$ milliseconds, \evo will wait $(60 000 / N) - X$ milliseconds before making a new HTTP request.

Although this parameter was originally introduced in \evo only for empirical studies, it was a necessity as well for a proof-of-concept study at  Volkswagen AG.
APIs are tested in a testing environment, mimicking production settings, including DoS prevention/mitigation mechanisms.

\subsection{Features Implemented in the Second Year}

\subsubsection{Example Objects}
In black-box testing, there is a need for testers to provide meaningful input data to help the fuzzers.
This is especially important when data is dependent on the content of databases, and there is no info about it in the schema.
As previously discussed, this input data can be provided by adding \texttt{example} and \texttt{examples} entries in the schema.
Typically, these would be string and number entries.

One feature that was not supported in \evo, but turned out to be quite important, is the handling of examples defined for objects.
Consider a body payload in a \texttt{POST} request, which could be represented with a JSON object.
Such object could have $n$ fields, e.g., $n=10$.
The schema can then define examples with instances of this object, instead of specifying specific values for each field individually.
This is useful for when there is the need of test data with specific relations between different fields.
Defining examples for individual fields such as \texttt{`foo: a,b,c'} and \texttt{`bar: d,e,f'} is different from specifying example object instances such as \texttt{`\{foo:a, bar:d\}'} and \texttt{`\{foo:b, bar:f\}'}, where specifying combinations of values are enforced.

One issue here is that, if there are $n=10$ fields, and the example is using only 2 fields (e.g., \texttt{foo} and \texttt{bar}), what to do with the other 8 fields?
If they are optional, we simply ignore them.
But, if they are required, then we must provide valid values.
We do this at random, while still satisfying the schema's constraints (e.g., regex on strings and min/max bounds on numbers).

From the point of view of the testers, they might want to make sure to test some specific combination of parameters, or there might be complex intra-parameter constraints that are hard to satisfy when generating data at random.
As such, the testers might want to specify some specific combination of values for some fields of the input objects, without though the effort of having to specify each single required field.
This can be left to the fuzzer to handle automatically.

\subsubsection{Multi-File Schemas}
To be able to know what to call in a REST API, an OpenAPI schema is required.
This can be specified as a location on the local file system, or as a URL to a remote server the schema can be downloaded from.

One important feature in OpenAPI is the use of \texttt{\$ref} links.
For example, if you have $n$ endpoints that take as input the same body payload structure, instead of copy\&paste such structure definition $n$ times, it can be done just once, and then having \texttt{\$ref} links pointing to it.
For example, a definition like \texttt{"\$ref": "\#/definitions/BodyDto"} would mean to check, in the current schema file, for the entry \texttt{definitions.BodyDto}, and use the information provided there.

As practically any non-trivial API schema uses \texttt{\$ref} links (as for example the open-source APIs collected in WFD~\cite{sahin_2025_wfc}), those are (or should be) supported by most fuzzers.
One major problem though, which we never encountered with open-source APIs, is that \texttt{\$ref} might point to external files.
In large organizations, where hundreds and thousands of APIs can be developed, re-using some data structure definitions among different APIs is not uncommon, as we experienced at Volkswagen AG.

The referenced schemas could be located on the same remote server or local file system using relative paths, e.g.,
\texttt{"\$ref": "../foo/bar.yaml\#/definitions/BodyDto"}
Or they could point to external services, e.g.,
\texttt{"\$ref": "http://localhost:8080/foo/bar.yaml\#/definitions/BodyDto"}.
Those URLs might even omit the communication protocol, like for example in
\texttt{"\$ref": "//localhost:8080/foo/bar.yaml\#/definitions/BodyDto"}.
In those cases, the protocol is inferred from the protocol used to download the current schema using the references.
In OpenAPI schemas there are quite a few edge cases like this to define \texttt{\$ref} links.\footnote{\url{https://swagger.io/docs/specification/v3\_0/using-ref/}}

To be able to fuzz some REST APIs at Volkswagen AG, we had to support all these kinds of external \texttt{\$ref} entries.
Those referenced schemas are automatically downloaded (users do not need to do anything manually).
If those schemas have themselves references to other schemas, those are downloaded and analyzed recursively.
This creates a graph of linked schemas, with possible cycles (which we handle).
Each individual schema is downloaded only once.

\subsubsection{Test Summaries as Comments}
When generating test cases for a non-trivial API, many test cases could be generated.
The actual code of a generated test could be quite long, as it might include several HTTP calls, each one having a large number of inputs (fields in body payloads and query parameters), and assertions on the fields of the returned responses.

There was a practical need, for each generated test, to have ``summaries'' of what a test does.
This can help to get a quick overview of what a test case is checking~\cite{panichella2016impact,djajadi2025using}.
We automatically generate test summaries in a deterministic way, by listing all the endpoints executed in a test, and what status code they return.
We report any found faults.
Faults are classified with WFC codes~\cite{sahin_2025_wfc}, with a short explanation directly added as comment on the HTTP calls that trigger the faults.
Also, for traceability, we also list which example values and links were used (if any).

\begin{figure}
\begin{lstlisting}[language=java]
/**
* Calls:
* 1 - (200) POST:/api/links/create
* 2 - (200) GET:/api/links/users/{name}/{code}
* Found 1 potential fault of type-code 101
* Followed 1 link:
*   200:LinkToGetUser
*/
@Test(timeout = 60000)
public void test_1_getOnUserReturnsMismatchResponseWithSchema() throws Exception {

    // Fault101. Received A Response From API With A Structure/Data That Is Not Matching Its Schema. Type: validation.response.body.schema.type. [Path '/errrors'] Instance type (null) does not match any allowed primitive type (allowed: ["string"])
    ValidatableResponse res_0 = given().accept("*/*")
            .post(baseUrlOfSut + "/api/links/create")
            .then()
            .statusCode(200);
    String link_0__data_id = res_0.extract().body().path("data.id").toString();
    String link_0__data_code = res_0.extract().body().path("data.code").toString();

    given().accept("*/*")
            .get(baseUrlOfSut + "/api/links/users/" + link_0__data_id + "/" + link_0__data_code + "?name=BAR")
            .then()
            .statusCode(200);
}
\end{lstlisting}
\caption{\label{fig:test-new}
Example of generated test like in Figure~\ref{fig:test} but with the latest version $4.0.0$ of \evo.
}
\end{figure}

Figure~\ref{fig:test-new} shows the updated version of the test shown in Figure~\ref{fig:test}, but this time generated with the latest version of \evo $4.0.0$ instead of $3.3.0$.
A summary is added as a code comment before the test is implemented.

\subsubsection{Test Naming}
Before version $4.0.0$, \evo generated tests that are named with a counter, like \texttt{test\_0} and \texttt{test\_2} (recall Figure~\ref{fig:test} and Figure~\ref{fig:token}).
This is not good for practitioners, as it increases the cognitive load needed to understand what the generated tests do.
In this regard, test generators based on LLM techniques scored better than \evo~\cite{poth2025technology}.

In the literature, in particular for unit test generation, different techniques have been developed to provide better test names~\cite{daka2015modeling,daka2017generating}.
There is a clear avenue for research on novel techniques tailored for REST API testing.
However, readability is not really something that can be objectively quantified (e.g., like achieved code coverage or detected faults).
It requires ``human studies'', i.e., to check if indeed a novel naming strategy is better than current practice you need test engineers to manually look at and evaluate those generated tests.
Unfortunately, user studies with human subjects in industry based on tool outputs are very risky, expensive and  tricky to carry out and then publish in the software engineering research community~\cite{davis2023s}, as we have unfortunately directly experienced for example in~\cite{zhang2025fuzzing} and in~\cite{garrett2025generating}.
For example, rejection comments such as the following are common:
``\emph{Industrial case studies, which are limited to a single company, can lead to bias}''.\footnote{Anonymous reviewer.}
As such, this type of studies is rare in the literature.
Apparently, the typical non-industrial case studies, which are limited to zero companies (i.e., all is done in the ``lab''), do not suffer of bias issues.

To improve the readability of the generated tests, as of version $4.0.0$ now \evo generates tests with meaningful names.
These names are generated with deterministic rules, based on the content of the tests (e.g., which endpoints are called, and what was returned in response by the API).
An example of this can be seen in the name \texttt{test\_1\_getOnUserReturnsMismatchResponseWithSchema} in Figure~\ref{fig:test-new}.
The details of this algorithm is presented in a separated work~\cite{garrett2025generating}.
The algorithm was then validated with a user study with 39 people recruited via LinkedIn,
comparing 8 different naming strategies (involving state-of-the-art LLMs as well) for 10 generated tests from 9 different open-source projects.
A further industrial validation at Volkswagen AG was carried out in~\cite{garrett2025generating}
to check the usefulness of this novel naming strategies on two industrial APIs.

This novel strategy improved the readability of the generated tests.
But, based on the received feedback, there is still more that can be done to improve them further~\cite{garrett2025generating}.

\subsubsection{DTO to Support Test Modifications}
An automated test case generator like \evo is useful for creating test suites for regression testing, as well as creating executable tests finding faults that can be used for debugging such issues.
For these goals, the generated tests can be used as they are, without the need to modify them manually.

Still, fuzzers are a not a replacement for professional test engineers.
After a fuzzing session, there might be still few edge cases and scenarios that need to be represented with manually written test cases. But those test cases do not need to be written from scratch.
Automatically generated test cases can be used as a ``starting-point'', and modified/extended to represent the missing testing scenarios.

\begin{figure}
\begin{lstlisting}[language=java]
given().accept("*/*")
       .contentType("application/json")
       .body(" { " +
             " \"name\": \"_EM_0_XYZ_\", " +
             " \"age\": 464, " +
             " \"address\": { " +
             " \"street\": \"4n7Qd86zLbDN1\", " +
             " \"country\": \"_EM_1_XYZ_\", " +
             " \"voted\": true " +
             " }, " +
             " \"key\": {} " +
             " } ")
       .post("${baseUrlOfSut}/object")
       .then()
       .statusCode(200)
       .assertThat()
       .contentType("text/plain")
       .body(containsString("OK"))
\end{lstlisting}
\caption{\label{fig:no-dto}
Example of HTTP call in which a JSON body payload is defined in the test as a string.
}
\end{figure}

To achieve such a goal, not only a generated test case must be \emph{readable}, but also it must be \emph{easy-to-modify}.
One major issue though in the context of testing REST APIs is the presence of JSON body payloads, typically sent as payload in \texttt{POST} and \texttt{PUT} requests.
Such payloads could have many fields of different types, including objects, recursively.
When JSON body payloads are represented as string in the generated tests (which is the default behavior in \evo), then modifications to such payloads have to be done as changes inside such strings.
Figure~\ref{fig:no-dto} shows a simple example with a HTTP \texttt{POST} request with a JSON payload represented with a string.
If a tester need to add a new field in that payload, they would need to consult the OpenAPI schema for that specific endpoint to see the list and exact spelling of each modified or added field to the payload.
A problem, though, is that a misspelled field by mistake might be ignored by the JSON parsing of the API, especially if the parameter is optional.
Not all JSON parsers are configured to do a strict validation of incoming objects, where unrecognized fields might be simply ignored with no warning.

\begin{figure}
\begin{lstlisting}[language=java]
val dto_Person_1 = Person()
dto_Person_1.name = "_EM_0_XYZ_"
dto_Person_1.age = 464
val dto_Address_1 = Address()
dto_Address_1.street = "4n7Qd86zLbDN1"
dto_Address_1.country = "_EM_1_XYZ_"
dto_Address_1.voted = true
dto_Person_1.address = dto_Address_1
val dto_SuperKey_1 = SuperKey()
dto_Person_1.key = dto_SuperKey_1

given().accept("*/*")
       .contentType("application/json")
       .body(dto_Person_1)
       .post("${baseUrlOfSut}/object")
       .then()
       .statusCode(200)
       .assertThat()
       .contentType("text/plain")
       .body(containsString("OK"))
\end{lstlisting}
\caption{\label{fig:with-dto}
Same example of HTTP call in Figure~\ref{fig:no-dto}.
However, here the JSON body payload is defined in the test with a DTO object.
}
\end{figure}

A solution for this problem is to create Data Transfer Objects (DTOs).
Besides generating test cases, \evo can generate for each body payload a DTO class representing its fields.
These DTO classes are then saved in a subpackage of the generated tests (e.g., under a package name \texttt{*.dto}).
Then, in the generated tests, each time a body payload is needed, an instance of these DTO classes is created and initialized with the right data.
Once instantiated, such DTO object is given as input to the HTTP library making the calls (e.g., RestAssured in our case), which, internally, will convert such DTO objet into a string representation.
Figure~\ref{fig:with-dto} shows the same test example for Figure~\ref{fig:no-dto}, but, this time, by using DTOs instead of raw strings.

If a test engineer needs to modify such a test case, they can leverage their IDEs and all their code-complete functionalities to easily modify those DTO instances, without the need to manually checking the schema.
This enforces no misspells in the selected fields.
However, this only works for statically typed languages such as Java and Kotlin.
Practitioners that use \evo to generate test suites in JavaScript or Python will not be able to use such DTO feature.

Generating the source code for DTO classes from OpenAPI schemas requires solving some technical challenges, like how to deal with \texttt{oneOf}, \texttt{anyOf} and  \texttt{allOf} constraints.
In our context, in the DTOs we enable the presence of all declared field names, even if presence of some of such combinations might result in a payload that is not valid according to the schema (e.g., in the case of \texttt{oneOf}).
When instantiating DTOs, we only consider syntactic validity, and not additional constraints expressed in the OpenAPI schema.
To ease their use, those constraints are added as code comments in the declarations of the fields and their setters/getters.

An important issue to deal with the creation of DTO code is how to distinguish between \texttt{undefined} and \texttt{null} in JSON.
The JSON objects \texttt{\{"x": null\}} and \texttt{\{\}} are different.
In the first object, the variable \texttt{x} is \texttt{null}.
In the second object, the variable \texttt{x} is \texttt{undefined}.
Unfortunately, in contrast to JavaScript, languages such as Java and Kotlin have no concept of \texttt{undefined}.
If a JSON string is naively marshalled into a DTO class such as \texttt{class DTO\{ String x;\}}, then there is no way to distinguish on whether the field was supposed to be  \texttt{null} or \texttt{undefined}.
This is a major issue, as such distinction can be critical for the correct operation of the APIs.
For example, in JSON Merge Patch RFC7386\footnote{\url{https://datatracker.ietf.org/doc/html/rfc7386}} the use of \texttt{undefined} is critical to leave fields untouched, as \texttt{null} is used to instruct their deletion.

A relative easy solution is to employ \texttt{Optional} Java fields, which are specially handled by Jackson (the JSON library used in RestAssured to marshall the DTO instances into strings).
For example, a DTO class could be \texttt{class DTO\{ Optional<String> x;\}}.
If the field \texttt{x} is left unset (i.e., default value \texttt{x=null}, and not an \texttt{Optional} instance containing the value \texttt{null}, like \texttt{Optional.ofNullable(null)}), then it is treated as \texttt{undefined} in the generated JSON (i.e., Jackson will generate \texttt{\{\}}).
Otherwise, it will instantiate with an \texttt{x} whose value depends on the content of the optional field, which can be  a string or \texttt{null}, e.g., \texttt{\{"x": null\}} and \texttt{\{"x": "foo"\}}.

The use of DTOs can simplify the \emph{modification} of the generated tests.
However, it does not necessarily mean that it improves its \emph{readability}.
This depends on the user's preferences.
As such, as most features in \evo~\cite{arcuri2023building}, the use of DTOs is configurable with a command-line option.

\subsubsection{Automated Data Cleanup}
As we have previously discussed, using existing data in the databases (if any is used by the tested API) would be beneficial for the fuzzing.
However, once a fuzzer starts to interact with existing data, there is no guarantee that the state of the database will be the same after the fuzzing has ended.

Creating new data (e.g., via \texttt{POST} requests) might not seem problematic, as long as the fuzzing does not create gigabytes or terabytes of data per hour.
However, it turned out to be a major showstopper for the testing of some APIs at Volkswagen AG.

Typically, when creating a new resource, this needs to be uniquely identified.
As all the resources created by all users all need to be unique, typically it is the API itself that creates the ids to guarantee their uniqueness.
This could be based on database sequence generators, or by using UUIDs.
When a user creates a new resource with for example a \texttt{POST} request with a body payload, such body payload will not have the id.
The server creates the resource and assigns a new unique id.
This information needs to be then sent back to the user (e.g., as part of the response of the \texttt{POST}), so they can interact with the newly created resource by using its id in following requests.

The problem happens when it is the user that decides the id, and sends such information as part of the resource creation operation.
An example is ``user registration'' where the id to identify the user is chosen by the user themselves.
If another user has already chosen the same user id, then the call to create the user will fail (e.g., the \texttt{POST} request might return a 400 HTTP status code).
This is problematic because it makes such tests unusable.
The first time a test is executed during the fuzzing process it might return a 201 (resource created), but, then, when run by the testers it will now return 400.
To be able to test the correct creation of a user, the user should not be registered yet.
Re-executing a test case will fail if the user id has not been changed.

To address this fundamental issue,
 we add explicit \texttt{DELETE} operations  after each test case execution, for each successful (e.g., status code in the 2xx family) create operation in it.
 However, in a black-box approach, determine which data is created, and which endpoint would result in deleting it, is not always simple.
 For example, not all APIs follow REST architectural guidelines on how to design hierarchical endpoints to manipulate resources in a predictable way.
We designed a simple pattern matching algorithm based on REST API guidelines to check, for each create operation (e.g., \texttt{POST} and \texttt{PUT}), what is the most likely \texttt{DELETE} operation for it (if any is defined in the OpenAPI schema).
For example, it is able to identify that a \texttt{DELETE:/user/\{id\}} is likely the correct delete operation for a \texttt{POST:/users}.
In this context, false positives are not an issue.
Worst case scenario, we are simply adding an unrelated \texttt{DELETE} operation that will simply return a 404 not found.

The handling of the automated clean up of resource creations, though, requires making new HTTP calls (i.e., \texttt{DELETE}).
This, however, could reduce the performance of the fuzzer, as those extra calls take time.
Is it better to have a fuzzer that makes more diverse calls?
Or a fuzzer that tries to clean-up after itself?
In our context, this latter option was preferred.
But it might not be the best option in all contexts.

\subsection{Features Summary}

\begin{result}
{\bf RQ2:} the most needed features are related on how users can provide extra guidance to the fuzzing, how to do authentication, how computational resources can be optimized and how to make the generated tests easier to read and understand.
\end{result}

\section{RQ3: Comparison With Existing Tests}
\label{sec:rq3}

Several of the enhancements introduced in \evo were related to its usability at industrial level.
Still, ultimately, what is important is how the generated tests can help the testers and QA engineers in their testing tasks.
As such, how these generated tests compare with the existing manually written tests is of major importance.
The closer in quality they are, the better the tool would be when addressing the testing of a new API for which no test exists yet.

To evaluate the actual benefits that \evo can provide to practitioners, we carried out an empirical study at the end of the first year in 2024 (Section~\ref{sub:first}).
This study was then repeated after the end of the second year in 2025 (Section~\ref{sub:second}).
Comparisons with the results created by 11 AI-test specialists from 4 different companies are then discussed in Section~\ref{sub:specialists}

\subsection{Experiments in First Year}
\label{sub:first}

Initial experiments done by tests engineers at Volkswagen AG with older versions of \evo (e.g., $1.6.1$) and with LLM techniques showed that manual tests are still significantly better~\cite{poth2025technology}.
In this paper, we repeated such experiments in 2024 with the latest version of \evo at that time (i.e., version $3.3.0$).
Furthermore, API schemas were enhanced with extra information, in particular regarding \texttt{examples} and \texttt{links}, as now these features are supported by \evo.

\begin{table}[!t]
    \centering
    \small
    \caption{Statistics of the two used industrial APIs in the experiments, including their name, lines of code (\#LOC),
    number of HTTP endpoints (\#Endpoints), and the number of the existing already written tests (\#Tests).
    }
    \label{tab:apis}
        \begin{tabular}{lrrr}\\
        \toprule
        Name & \#LOC & \#Endpoints & \#Tests \\
        \midrule
        \texttt{auth-service} & 5750 & 3 & 16 \\
        \texttt{user-service} & 8591 & 21 & 56 \\
        \bottomrule
        \end{tabular}
\end{table}

The experiments involve two APIs developed at Volkswagen AG, chosen by its test engineers as initial case study for evaluating the potential benefits of automated test generation, as part of the innovation activities of QiNET, which is part of TQA.
Table~\ref{tab:apis} shows some statistics of these two APIs.
These APIs have been curated, thorough test suites, covering all different aspects of importance deemed by the test engineers at Volkswagen AG.
Those reference tests are hence used as ``gold standard'' for these 2 APIs in our empirical analyses.

\begin{table*}[!t]
    \centering
    \small
    \caption{
    Evaluation, in the first year, of the tests generated by \evo compared to the existing manually written tests.
    Out of the \emph{total} number of generated tests,
    we report how many tests are \emph{accepted} (i.e., covering same scenarios as in manual tests, without need to be modified),
    how many needed to be manually \emph{modified} to be usable,
    how many were \emph{removed},
    and how many tested \emph{new} scenarios that are not covered in the manual tests.
    We also report how many test scenarios from the handcrafted tests are \emph{missing} in these generated tests.
    }
    \label{tab:results}
        \begin{tabular}{ l r| rrrr |r }\\
        \toprule
        Name  & total & accepted & modified & removed & new & missing  \\
        \midrule
        \texttt{auth-service} & 8  & 8  & 0 & 0 & 0 & 8 \\
        \texttt{user-service} & 55 & 46 & 0 & 2 & 7 & 8 \\
        \bottomrule
        \end{tabular}
\end{table*}

On these APIs, we ran \evo three times,
for a time budget of 10 minutes.
The outcome variation was between 55 and 60 test cases for the \texttt{user-service}.
The outcome was always 8 test cases for the \texttt{auth-service}.
However, a longer search time does not seem to improve the outcome significantly, as still in the last minutes of the 10 minute search no additional testing target was covered.
For each API, we took the ``worst case'' of the generated test suites, with the minimum number of test-cases out of the three runs, for a conservative evaluation of the results.
Table~\ref{tab:results} shows the results of this manual analysis
 after verification by the engineers against the handcrafted reference test-suites referenced in Table~\ref{tab:apis}.

This comparison was manually done by test-engineers, with more than 10 years of experience, within an industrial setting.
In particular, we are interested in checking how many generated tests could be kept (i.e., they
are evaluated as ``good''), how many need to be modified (e.g., for LLM-based generated tests in previous work~\cite{poth2025technology}, most of them needed to be modified to remove hallucinations related to generated assertions on the returned responses), how
many need to be removed (i.e., they are redundant or simply
considered not useful), and how many are missing to reach
the same quality level of the reference implementation (i.e.,
the already existing manually written tests).
Also, we are interested in what was newly discovered and added to the test-suite by \evo.

As we can see in Table~\ref{tab:results}, many generated tests are useful.
Overall, the generated tests save time and effort for the engineers.
However, the generated tests are not at a level that they are directly usable as test-suites.
They still need validation and enhancement by engineers to reach a sufficient quality level for a usable test-suite.
But, the automatically generated test-suites also add tests which would usually not be written by engineers, as too much work for less ``value''.
Here, the AI-facilitation offers a chance to improve the test-suite quality beyond the existing handcrafted test cases.
Notable test-suite enhancements are the typical test set of the lower and upper boundaries and a random data set.
The engineers often do not define all potential variants.
This results in that the \evo and a test-engineer ``pair'' can deliver in less time a better test-suite than only a handcrafted one would be under real-world economical constraints.
Still, a major impediment for future improvement is the verification effort of the generated tests.
Here, activities are needed on the \evo test-case generator to make the verification less time and effort consuming, e.g., by improving the readability of the generated tests
and provide some kind of traceability to the used OpenAPI schema components.

\subsection{Experiments in Second Year}
\label{sub:second}

\begin{table*}[!t]
    \centering
    \small
    \caption{
    Evaluation, in the second year, of the tests generated by \evo.
    Out of the \emph{total} number of generated tests,
    we report how many tests are \emph{accepted} (i.e., covering same scenarios as in manual tests, without need to be modified),
    how many needed to be manually \emph{modified} to be usable,
    how many were \emph{removed},
    and how many tested \emph{new} scenarios that are not covered in the manual tests.
    We also report how many test scenarios from the handcrafted tests are \emph{missing} in these generated tests.
    The value ``N/A'' missing that the data could not be collected or evaluated.
    }
    \label{tab:results-second}
        \begin{tabular}{ l r| rrrr |r }\\
        \toprule
        Name                  & total & accepted & modified & removed & new & missing  \\
        \midrule
        \texttt{auth-service} & 22  & 12  & 0 & 1 & 9   & 4 \\
        \texttt{user-service} & 62  & 48  & 0 & 4 & 10  & 8 \\
        \texttt{P1}           & 11  & 11  & 0 & 0 & N/A & 3 \\
        \bottomrule
        \end{tabular}
\end{table*}

At the beginning of 2026, we repeated the same experiment, with the latest version of \evo at that time (i.e., $5.0.2$).
The 2 previous APIs were used exactly as they were, with no modification to their source code (i.e., the same exact versions were used).
Besides the 2 APIs used in the first year, we also included 2 other APIs from different departments at Volkswagen AG.
Due to the black-box nature of this kind of testing, the test engineers were not able to provide detailed information about these APIs developed in different departments (e.g., LOCs).
As these APIs did not have a thorough test suite that could be reliably used as gold standard, some data could not be collected.
Furthermore, due to confidentiality reasons, we cannot name those new APIs.
In this text, we refer to them with the labels \texttt{P1} (which is part of a vehicle related API) and \texttt{P2} (which is part of an IT platform).
Table~\ref{tab:results-second} shows the results of this second study.

 The evaluation for these new APIs utilized a basic OpenAPI schema that did not include extensive examples or supplementary semantic information.
 Such scenarios are common in industry, where high-quality documentation is frequently absent.
 Due to the absence of a ground truth, data for the \emph{new} category could not be collected.
 Tests identified as \emph{missing} involved the use of incorrect access data, incorrect content type, and the omission of a refresh request.
 For \texttt{P1},  \evo successfully generated 11 out of 14 tests, corresponding to a success rate of 79\%.
 The evaluation for \texttt{P2} was conducted under the same specification context.
Despite these differences, the automated approach still provides significant advantages for test engineers in real-world scenarios.
Most of the common scenarios are properly handled, with only few cases that are \emph{missing} and could not be automatically generated.
Compared to the results of the first year, \emph{new} scenarios are covered, and the number of \emph{missing} has been reduced.


Further extending the OpenAPI schemas to incorporate additional details and semantics of the API, aiming at getting better results with a fuzzer, becomes increasingly challenging.
Firstly, the OpenAPI specification has inherent limitations, making it difficult to represent higher-level semantics such as workflows within an API schema.
Secondly, the effort required to include these details involves a trade-off: whether it is worthwhile to invest significant time in augmenting the specification when missing or modified features can be addressed directly by modifying/extending the generated test suites.

These real-world constraints provide new research directions for \evo development, to focus resources toward features that can be readily utilized by test engineers, such as lifecycle management of test suites and facilitating the re-validation of new test suite versions against their last validated release (this will be covered in more details in Section~\ref{sec:rq4}).
For instance, as demonstrated by the \texttt{user-service} in Table~\ref{tab:results} and Table~\ref{tab:results-second}, incorporating lifecycle history narrows the gap to only seven tests, compared to validating the entire test suite version from scratch with all the 62 tests.
The value oriented focus from a lifecycle perspective should be to reduce 90\% of the validation effort for the test suite, rather than striving to optimize the remaining 10\% of test case generation.
The missing test cases could be manually added during the initial validation of the first test suite release, and then could be readily incorporated into subsequent versions.

\subsection{Comparison with AI-test Specialists}
\label{sub:specialists}

\begin{figure}
\includegraphics[width=0.95\textwidth]{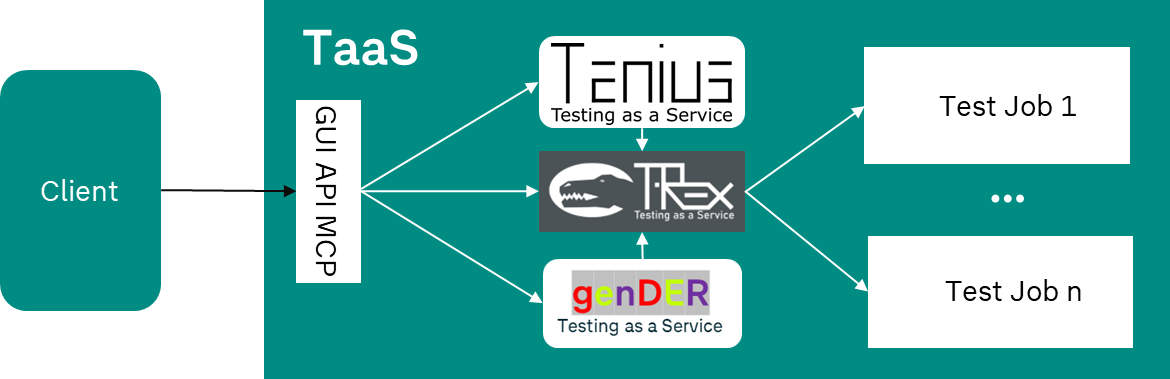}
\caption{ \label{fig:taas}
High-level TaaS architecture and three core services. Tanius offers \evo as a test-case generator.
}
\end{figure}

Test engineers at Volkswagen AG can easily use \evo in their work, as it is integrated into the Testing as a Service (TaaS) system Tanius, internally used for test case generation needs~\cite{poth2025technology}.
Figure~\ref{fig:taas} shows on a high-level perspective that TaaS uses different components to execute test jobs.
Tanius includes \evo to generate test-cases which are executed by T-Rex as test jobs.
The genDER component is used to design the test-job runtime~\cite{poth2025generative}.
However, every tool has its own limitations and constraints.
To better assess the performance of \evo compared to the established approaches in \emph{industry}, an open experimental setup was set up.

The setup involved, on a voluntary basis, several existing framework contract partners from IT consulting, service, and solution companies.
Participants were tasked with addressing challenges using any methods or tools they deemed appropriate.
There were no restrictions, allowing experts to employ whatever strategies they believed would be the most effective.

Four companies took part in the challenge,
each forming expert teams (with a maximum of three members) specializing in AI-driven test case generation, for a total of 11 participants.
These teams significantly outperformed ``typical'' testing professionals in leveraging AI for test engineering tasks, as the companies sent their best and most experienced experts to solve the challenge.
To avoid conflicts between the teams, each team performed the challenge at different times, on Volkswagen AG's premises.
The challenge took place between December 2025 and January 2026, and was timeboxed to three hours.

One of the tasks involved replicating the exercise previously conducted by \evo (recall Table~\ref{tab:results-second})
for the  \texttt{auth-service} and \texttt{user-service} APIs.
As reference, 91 tests are needed for a full test-coverage ($16+56=72$ for the original reference, plus the $9+10$ \emph{new} scenarios identified by \evo), where 79 tests were correctly generated by \evo (i.e., $4+12$ needed test scenarios are \emph{missing}).
All participants were provided with the same OpenAPI schema and API as the test object at the start of the challenge.

The preferred tools for addressing the challenge included IDEs such as Visual Studio Code, supplemented by large language models (LLMs). Some teams relied on a single LLM, while others used multiple LLMs depending on the specific task.
The chosen LLMs  included MS/Github Copilot, GPT-5, Gemini 3, Claude Opus 4, and Sonnet 4.
Although participants had the option to use proprietary AI tools, none chose to do so.
Only one team decided not to use a LLM.
They relayed on their own methodology to derive test-suites from OpenAPI schemas.

Approaches varied: some teams prioritized the quality of generated test cases over quantity, while others focused on producing a high number of test cases.
Despite these differences, all participants succeeded in generating valuable outputs which were presented at the end of the challenge.
 Notably, teams that emphasized framework organization and maintainability produced robust testing infrastructures/setups, but did not achieve high coverage within the three-hour window.
 Conversely, the team that prioritized quantity generated over 150 tests, but did not validate their ``quality''.
 Random sample checks of these generated tests revealed that many tests returned a ``passed'' status for both 2xx and 4xx HTTP responses (e.g., with statements like \texttt{assertTrue(code >=200 \&\& code <= 499)}, which is in most cases not particularly useful without refinement to one of the two response classes.
 This shows how important it is to manually validate each generated test, especially when using techniques such as LLMs.

Overall, the teams were able to delegate tasks to LLMs and achieved time savings.
One team started with 2 persons (one engineer and one coach) and delivered similar outcome like the 3 person team (three engineers, as all other teams) without LLM facilitation.
However, no team managed to achieve both high quality and high quantity in test case generation as effectively as \evo.
This was based by a manual review of all generates tests against the reference solution test suites for the challenge.

The key observation from this challenge is that, even highly skilled test engineers, given complete freedom in methodology and tool selection, in three hours could not surpass the performance of \evo run for few minutes
and use the rest of the 3 hours to validate the generated tests and do some enhancement on the test-suite.
This performance is now accessible to ``typical'' test engineers at Volkswagen AG via TaaS.

\begin{result}
{\bf RQ3:} The tests generated by \evo are of practical usefulness in industrial contexts.
Compared to existing manually written tests, these generated tests can cover many similar scenarios, as well as new ones.
However, few important test scenarios are missing in the generated tests, which makes them not a full substitute for the manual tests.
Compared to what produced by AI-test specialists with complete freedom of using what they wanted, \evo gives significantly better results in a fraction of the time  in the given use-case setup.
\end{result}

\section{RQ4: Open Challenges}
\label{sec:rq4}

\evo is currently used by some test engineers at Volkswagen AG.
Several challenges still need to be addressed before it can be used more widely inside Volkswagen AG.
Improving fuzzing algorithms to increase coverage and fault detection is always beneficial.
However, we will not discuss those latter here.
We will rather focus on the challenges and features directly pointed out and requested by the test engineers involved in this work.
Note that these challenges and features have nothing specific to Volkswagen.

{\bf Domain Knowledge}.
There is a need to enhance and guide fuzzers with domain knowledge.
This can be complex, representing business requirements and specific chains of API calls with specific database states.
In a black-box testing context, it would likely be infeasible to automatically discover such information within reasonable time.
Using \texttt{examples} and \texttt{links} in the OpenAPI schemas is a useful first step, but it is not sufficient.
How to best provide information on such domain knowledge, and how to make sure that such information can be properly ``interpreted'' by the fuzzers, are open research questions.
Furthermore, the \emph{traceablity} from the generated tests to this provided domain knowledge is crucial to verify sufficient correctness and completeness of the generated test suites.

At the time of writing, the OpenAPI Initiative~\footnote{\url{https://www.openapis.org/}}
(OAI) consortium (a Linux Foundation Collaborative Project) has recently created and released the new
\emph{Arazzo} Specification:
\emph{``The Arazzo Specification defines a standard, programming language-agnostic mechanism to express sequences of calls and articulate the dependencies between them to achieve a particular outcome, or set of outcomes, when dealing with API descriptions (such as OpenAPI descriptions)''}.\footnote{\url{https://www.openapis.org/arazzo-specification}}
If testers provide sequence of call workflows using Arazzo, those could be used and exploited by fuzzers such as \evo to generate test cases using those specific workflows as starting point templates.

{\bf Seeding Tests}.
Fuzzers are not meant to replace human software testers.
For sure, at least not as long as the results of fuzzers are worse than test suites manually crafted  by experienced test engineers~\cite{poth2025technology}.
Especially when dealing with existing APIs, in professional environments there might be existing end-to-end tests for those APIs.
This could be written in programming languages such as Python and Java, using specialized libraries to make HTTP calls.
Or they could be written in a domain-specific-languages (DSL) for specific API testing tools, such as for example
Postman,\footnote{\url{https://www.postman.com}}
Insomnia,\footnote{\url{https://insomnia.rest}}
Hoppscotch\footnote{\url{https://hoppscotch.io}}
 and
Bruno.\footnote{\url{https://www.usebruno.com/}}
Using these existing tests (if any) as \emph{seeds} for the fuzzers could boost their effectiveness, providing  interesting starting test cases to fuzz.
Seeding existing tests is a known technique to reuse existing domain knowledge, like for example done in search-based unit test generation with tools such as EvoSuite~\cite{rojas2016seeding}.
As such, seeding end-to-end tests for REST APIs is an important feature.

In the past, in \evo there was a first attempt, i.e., a proof-of-concept, to seed tests written for the Postman tool~\cite{martin2021black}.
However, this so called test ``carving'' process was a fool's errand.
Unfortunately, there are just too many ways in industry to write end-to-end tests for REST APIs.
Trying to support all of them is infeasible.
Even if one concentrate on a single type, e.g., JUnit tests, there are simply too many ways to write test cases and so many different libraries used to make HTTP calls.
 Implementing a test carver that can handle all these different cases is simply not a viable option.

 Fortunately, though, in the case of REST APIs, a feasible alternative could be to use a HTTP proxy.
 Regardless of how the existing tests are written (e.g., in Python or in the DSL used by Bruno), they make HTTP calls.
All these HTTP calls can be intercepted and analyzed with a proxy, regardless of how the tests were written.
At a high level, this would be similar to how black-box coverage tools such as Restats work~\cite{corradini2021restats}.
Then, from these HTTP logs, the proxy can output a format representing the executed tests that can be directly used by the chosen fuzzer (e.g., \evo in our case).

One challenge here, though, is how to deal with dynamic data.
In other words, when some output of a call is used as input in a following call.
For example, recall the cases of
Figure~\ref{fig:token}.
All these cases would need to be detected and handled by the proxy.
If not, if the input data is used as it is (e.g., with hardcoded authentication tokens), then the seeded tests could become flaky.

One advantage of test seeding is that it can enable \emph{test enhancement} as well.
Given an output test suite generated by \evo, testers could modify it to improve it.
These improvements will not be lost if then these modified test suites can be fed back to the fuzzer as new seeds.

Another alternative approach to use an HTTP proxy would be to use an LLM to translate the existing manually written test cases (regardless of the programming language they are written in) into a format that a fuzzer like \evo can directly handle.
Regardless of the costs and constraints of using LLMs, this would be just a ``one-time'' endeavour when translating previously existing tests.
However, if this is set up to enable test enhancement as well during the whole lifecycle of testing the APIs, the costs of using the LLM would need to be taken into account.

{\bf Test Case Reuse}.
When academics run experiments with different fuzzers, or different configurations of the same fuzzer, those are run for a certain amount of time.
This could be for example 10 minutes, 1 hour or 24 hours.
But this is not really close to the development cycles in industry.
Usually, a REST API is not implemented in one day, tested once, and then be done with it.
It might takes weeks, months or years to develop and maintain/update an API.
Each day, any new code change in the API could require and warrant a new fuzzing session.
Therefore,  throughout the lifespan of an API, it might require to be fuzzed hundreds of times.

With the passing of time, each new introduced code change might invalidate any existing regression test.
Still, fuzzing from scratch each day would be  inefficient.
It would likely be beneficial to re-use the generated tests from the latest run like ``seeds''.
This would create a chain of seeding, starting from the first run.
However, as the semantics of the API and its OpenAPI schema might change through time, there is the need to make sure to do not crash the fuzzer if any previous test case is no longer valid (e.g., referring to an endpoint or parameter that no longer exist).

Conceptually, this is very similar to the previously discussed seeding from existing tests.
Two differences though:
(1) as already mentioned, some tests might be obsolete, or referring to no longer existing parameters. A fuzzer should not crash in those cases;
(2) besides the output type chosen by the user (e.g., JUnit 5 in Kotlin), a copy of the tests can be saved as well in any internal representation used by the fuzzer, as only the fuzzer needs to read those copied tests.
This latter point should make this feature easier to implement than, for example, a proxy used to record and analyze the execution of existing tests on-the-fly.

This concept of seeding from previous executions has already been investigated in the literature, for example in the context of unit test generation in Continuous Integration~\cite{campos2014continuous}.
From a research standpoint, there would not be much novelty here.
Still, this would be a critical feature to have for a successful technology transfer from academic research to industrial practice.

{\bf Database Handling}.
Once a fuzzer starts to interact with existing data in the testing environment databases, there is no guarantee that the state of the databases will be the same after the fuzzing has ended.
If other manual test cases rely on specific existing data, this can be a problem.
Furthermore, interacting with existing data might create unwanted dependencies between different tests.
Rollbacking the database to a previous snapshot after each fuzzing session is technically possible, but not ideal.

 The case of deleting or changing existing data is problematic.
 One approach could be to do not directly interact with such data, but rather read it (e.g., with \texttt{GET}) and then make copies of it (e.g., create new resources with same data using operations such as \texttt{POST}).
 Then, in each test that creates new data, \texttt{DELETE} operations can be automatically added.

Handling these issues might require making new HTTP calls (e.g., \texttt{DELETE}), using different strategies.
This, however, could reduce the performance of the fuzzer, as those extra calls take time.
How to best handle this issue is an important research problem.

{\bf AI Hype}.
Currently, as of 2026, there is a lot of hype about what Generative AI can do in industry.
Several fields have experienced disruptive innovations thanks to Generative AI.
As for everything, there are always strengths and weaknesses in any new approach.
There is a lot of potential for synergies between established techniques such as Search-Based Software Testing
and Foundation Models~\cite{fraser2025retrospective,sartaj2025searchbasedsoftwareengineeringai}.
Therefore, how these novel techniques can be best used in software testing is an ongoing research investigation~\cite{augusto2025large}.
The use of AI-based techniques was a decisive factor to choose \evo (recall Section~\ref{sec:rq1}).
However, further AI techniques could be exploited to achieve better results.

In the case of testing REST APIs, there has been some work in leveraging Large Language Models (LLM), like for example to analyze possible parameter constraints expressed in natural language documentation~\cite{kim2024leveraging}, or to sample possible valid inputs based on their name~\cite{corradini2024deeprest}.
Integrating this type of techniques inside \evo, or combining other LLM approaches with it,  is an important feature, based on the internal long-term vision for strategic focus in large enterprises such as Volkswagen AG.

{\bf Low-Code Testing}
Typically, users of a fuzzer would be either software developers (e.g., when doing white-box testing) or software testers (e.g., when doing black-box testing).
Still, domain experts with no programming knowledge might be involved when testing and discussing the results of the testing of   APIs.
This is the case at Volkswagen AG, and in many other enterprises as well.
Those domain experts might not be able to fully understand test code generated in programming languages such as Python and Java.

To enhance the participation of these domain experts in the testing process, there is the need of some \emph{low-code} solution~\cite{khorram2020challenges}.
Fuzzers could generate tests that are easier to understand for people with no programming knowledge.
The use of test summaries in natural language is a first step forward in this direction.
Interactive web applications could be developed in which test results of the fuzzers are only displayed in natural language with graphical interface support.
The actual test code could be hidden by default, where then it could be displayed (e.g., via a web-toggle) if needed by testers or programmers.

\begin{result}
{\bf RQ4}: how to best exploiting existing tests, as well as tests generated from previous runs of the fuzzer, is of paramount importance.
How to best name the generated test cases, and how to deal with databases, are also important challenges to investigate.
Furthermore, innovations from novel AI and low-code techniques are highly sought in industry.
\end{result}

\section{RQ5: Beyond Volkswagen AG}
\label{sec:rq5}

What presented in this paper is based on the experience in the last two years between the development team of \evo and some test engineers at Volkswagen AG.
However, \evo is used also in other enterprises, such as Meituan~\cite{zhang2023rpc,zhang2024seeding,zhang2025fuzzing}.
In this latter case though, the main interests are more focus on the white-box testing of RPC APIs (instead of black-box testing of REST APIs).

At the time of writing,  \evo has been open-source for more than nine years, with thousands of downloads.
Several enterprises around the world have been using it.
It has not been uncommon that developers and testers in industry that have used \evo contacted the \evo's team with ``feature-requests''.
This included for example supporting multi-schemas in database handling,\footnote{\url{https://github.com/WebFuzzing/EvoMaster/issues/806}}
as well as trying to provide mechanisms to handle flakiness in the generated tests.\footnote{\url{https://github.com/WebFuzzing/EvoMaster/issues/1278}}

When feature requests are easy to implement, supporting the open-source community is easy.
Any implemented feature-request makes the tool better.
However, there have been cases in which some feature-requests would had been quite complex to support.
Without an academia-industry   co-operation (where experiments can be reported on industrial systems), or if the feature-requests would not be relevant for the open-source APIs we use for experimentation (e.g., WFD~\cite{sahin_2025_wfc}), it can become hard to justify the time and resources spent in implementing such feature-requests if there is no direct academic benefit out of it (e.g., a resulting publishable scientific article).

Usually, ``we'' (i.e., the \evo's team, as Volkswagen has no relation to what discussed in this section) tries to establish working relations with practitioners that use \evo in industry, to do the kind of study reported in this paper.
But, several practitioners either have no time nor interest in such kind of work.
Or they might not have any authority to establish such kind of   co-operation (e.g., testers/developers without the backing of higher management).
In this latter case, though, we have experienced a quite interesting case.
Once, we received a quite complex feature-request related to RSA (Rivest–Shamir–Adleman) encryption of input test data.\footnote{\url{https://github.com/WebFuzzing/EvoMaster/issues/1154}}
None of the open-source APIs in WFD~\cite{sahin_2025_wfc} needs this, and it seems not relevant to neither Volkswagen AG nor Meituan.
For undisclosed reasons, no academia-industry   co-operation could be established with the user asking for such feature-request.

The problem to solve was as following:
\begin{itemize}
\item First a client creates a random AES (Advanced Encryption Standard) key, which will be stored in a \texttt{key} field in the requests toward the API.
\item The AES \texttt{key} needs to be encrypted using the \emph{public} RSA key of the API.
\item All the information sent by the client (e.g., all different kinds of data needed to be sent to use the different functionalities of the REST API) needs to be encrypted using the AES \texttt{key}, and saved as a string into a single \texttt{data} field in the requests.
\item  Once the \texttt{key} and \texttt{data} fields are computed, they are combined, and a new \texttt{sign} field is created by encrypting such combination with the client's \emph{private} RSA key.
\end{itemize}

When the API receives a body payload with these \texttt{key}, \texttt{data} and \texttt{sign} fields, it will do the following:
\begin{itemize}
\item Use the \emph{public} RSA key of the client to verify if \texttt{sign} is matching the content of \texttt{key} and \texttt{data} (and so the request was actually made by the client and not someone else).
\item Use its own \emph{private} RSA key to decrypt the AES \texttt{key} (so only the API can decode the message from the client).
\item Use the decrypted AES key to decrypt of the content of the \texttt{data} field.
\item Use such decrypted data as input to the invoked REST API endpoint.
\item If any of the previous checks failed, return an error response to the client.
\end{itemize}

An OpenAPI schema might have information about the entries \texttt{key}, \texttt{data} and \texttt{sign}.
However, without any further information provided by the user, there is no chance that a random string would correctly match data encrypted with RSA and AES cyphers.
All calls made by a fuzzer toward such API would  fail (e.g., 400 HTTP status).

The support to enable a fuzzer to deal with such kind of API would be a complex feature to implement.
If there is no available API to experiment with, and no access to the original industrial API is possible, then there would be no much point (from an academic perspective) in implementing such feature.
Furthermore, solutions that are ad-hoc for specific APIs might have limited scientific value.

In this case, though, it seems that this kind of RSA/AES encryption of data is not uncommon among finance-related APIs.
Still, the lack of accessible APIs to experiment with would be an academic showstopper.
In this particular case, this problem was solved by the user that made the feature-request.
They implemented an \emph{artificial} API to reproduce this kind of communications, and made it open-source.
We then could use this API to try out how to enable \evo to deal with this kind of APIs.

Under these conditions, to improve the applicability of \evo among enterprises, we agreed to try to solve this issue.
However, any derived solution should be general, and not tailored to any specific API.

To solve this problem, in \evo we introduced the concept of \emph{derived} parameters.
When there is a body payload as input to the requests (e.g., \texttt{POST} and \texttt{PUT}), the fuzzer will choose the input parameters (i.e., the fields in the object payload) as usual.
The user has the option to specify if some parameters are \emph{derived}, based on names.
If a body payload has a field that is marked as \emph{derived} based on this name matching, then a transformation function is applied.
Such a function takes as input the name of the derived field (with as well the name of thr endpoint), and the whole body payload.
It then returns the transformed value.
\evo will then use this transformed value for the derived parameter before making the HTTP call.

As transformations might be chained (e.g., recall the example of \texttt{sign} which needs to be derived from the transformed \texttt{key} and \texttt{data}), there is the need to express an ordering of transformations.
\evo can now automatically apply those transformations, in the right order, creating valid HTTP requests toward this kind of APIs.
Still, such transformation functions must be provided by the user.
As those functions can be arbitrarily complex (e.g., RSA encryption with specific cypher versions), they need to be provided with code (and not just configuration files like YAML or JSON).

\begin{figure}
\begin{lstlisting}[language=java]
@Override
public String deriveObjectParameterData(String paramName, String jsonObject, String endpointPath) throws Exception {  ©\label{line:derived:function}©

  if(paramName.equals("sign")){
    CommonReq req = JSONObject.parseObject(jsonObject, CommonReq.class);
    String signText = req.signText();
    return CryptoUtil.sign(signText, DemoController.OTHER_PARTY_PRIVATE_KEY);
  }

  if(paramName.equals("key")){
    return CryptoUtil.encryptByPublicKey(aesKey, DemoController.YOUR_PUBLIC_KEY);
  }

  if(paramName.equals("data")){
    CommonReq<BindCardReq> req = JSON.parseObject(jsonObject, new TypeReference<CommonReq<BindCardReq>>() {});
    return CryptoUtil.encrypt(JSONObject.toJSONString(req.getBizData()), aesKey);
  }

  throw new IllegalArgumentException("Unrecognized parameter: " + paramName);
}

@Override
public ProblemInfo getProblemInfo() {
  return new RestProblem(                                      ©\label{line:derived:rest}©
     "http://localhost:" + getSutPort() + "/v3/api-docs",    ©\label{line:derived:schema}©
     null
  ).withDerivedParams(Arrays.asList(                           ©\label{line:derived:settings}©
     new RestDerivedParam("key", DerivedParamContext.BODY_PAYLOAD, null, 0),
     new RestDerivedParam("data", DerivedParamContext.BODY_PAYLOAD, null, 0),
     new RestDerivedParam("sign", DerivedParamContext.BODY_PAYLOAD, null, 1)
  ));
}
\end{lstlisting}
\caption{\label{fig:derived}
Code excerpt for driver class in which derived parameters are configured.
}
\end{figure}

To solve this issue, currently we enabled the handling of derived parameters only for white-box testing, where users anyway need to write \emph{driver classes} (e.g., in Java or Kotlin) to tell how to start/stop/reset the APIs~\cite{arcuri2019restful}.
These transformation functions will be implemented in these drivers.
Then, the driver needs to tell \evo which parameters are marked as derived.
Figure~\ref{fig:derived} shows the two functions that in the driver classes need to be modified/implemented.

The function \texttt{getProblemInfo()} is the major entry point for the \evo drivers.
In this example, it specifies on Line~\ref{line:derived:rest} that we are dealing with a REST API  (and not a GraphQL or RPC one),
where the URL of where its OpenAPI schema can be found is specified on Line~\ref{line:derived:schema}.
The new settings to specify derived parameters are set on Line~\ref{line:derived:settings}.
Each parameter has four properties:
its name,
where it is located (currently we support only body payloads, but in the future we could consider URL query parameters as well),
the endpoints in which the transformation should be applied (\texttt{null} means any endpoint in which the parameter is defined),
and the transformation order (in this case, \texttt{sign} should be computed last, as depending on the transformation of the other derived parameters).
The implementation of the actual transformation is done inside the \texttt{deriveObjectParameterData()} function on Line~\ref{line:derived:function}.
This functon is then going to automatically be called by \evo to compute the derived parameters, when needed.

With this solution, we can fully handle the artificial example provided by the practitioner that requested this feature.
Our solution is general, and can be applied to any kind of API in which some parameters must be derived with complex rules that cannot be expressed with OpenAPI schemas.
Still, such solution requires some manual configuration, as shown in Figure~\ref{fig:derived}.
However, this solution seems practical (based on the feedback we received).
When direct academia-industry   co-operations are not possible, practitioners providing artificial examples to replicate the problem they face is an important venue to explore.

\begin{result}
{\bf RQ5}:  Feedback from industry is essential.
Many issues will be common among enterprises, but others might be specific to particular market domains.
The need to support derived parameters (e.g. for handling RSA/AES encryption) is such one example.
Engaging with the users community is highly beneficial to spot new, important research opportunities.
\end{result}

\section{Threats To Validity}
\label{sec:threats}

This industrial report shares our lessons learned in the enhancement of a specific tool (i.e., \evo) applied at a real-world enterprise (i.e., Volkswagen AG).
Threats to external validity include how these lessons learnt could be applicable to other tools and enterprises.
Most of the discussed enhancements are rather general, and likely any fuzzer that wants to be applicable to industrial APIs would need to implement the same or similar features.
Furthermore, there is nothing specific in the analyzed APIs that is limited to Volkswagen AG.
From a technical perspective, those APIs share many high-level characteristics with existing open-source APIs, such as the ones collected in WFD~\cite{sahin_2025_wfc} used for evaluating fuzzing techniques.
As such, it seems feasible that our results could generalize in other industrial contexts and tools, although this is not possible to know for certain without further studies.

The identified challenges and research priorities are based on the feedback from some test engineers in the Group IT of Volkswagen AG.
Even within the same organization, other test engineers might have different opinions.
Therefore, we cannot say for sure that what identified are indeed the most important challenges and priorities for fuzzing REST APIs in industry.
Regardless, those are important problems that need to be addressed.

Regarding the empirical study, threats to internal validity might come from faults in our new feature implementations.
For example, faults in the implementation of handling \texttt{examples} and \texttt{links} features might have negatively impacted our results.
As \evo is open-source on GitHub, with new releases automatically uploaded to Zenodo, anyone can review its code.

Another potential threat is that, as the generated test suites were manually evaluated with the existing test suites, such manual process might have been affected by human mistakes.
Unfortunately, due to confidentiality and intellectual property constraints, we cannot provide a replication package for this industrial study.

\section{Conclusions}
\label{sec:conclusions}

Automated testing of REST APIs is a topic that has attracted a lot of interest from the research community in the last few years~\cite{golmohammadi2023testing}.
A reason for this is that the verification of REST APIs is of paramount importance in industry, especially in large enterprises such as Volkswagen AG.
However, no experience report of introducing these novel techniques in industry has been presented in the academic literature so far~\cite{golmohammadi2023testing}, till now.

In this paper, we have reported our experience in the technology transfer from academic results (e.g., our \evo search-based fuzzer) to industrial practice (e.g., at Volkswagen AG).
We have discussed which technological and scientific challenges we have faced in making \evo more usable for practitioners in industry.
Practical solutions have been presented and evaluated.
Still, there are several challenges that need to be addressed, which we summarized in this paper.

The challenges reported in this paper do not seem to be specific to \evo or Volkswagen.
Likely, any other fuzzer or enterprise will face similar issues.
As such, this industry report can provide useful information for other researchers on fuzzing techniques that want to see them applied in industry practice.

Experiments on industrial APIs, and an empirical study with 11 AI-test specialists from 4 different companies, show that \evo can give better results than the current state-of-the-art in industrial practice.

The fuzzer \evo is open-source, and freely available on GitHub:
\url{https://github.com/WebFuzzing/EvoMaster}

\section*{Acknowledgments}
The \evo's team that contributed to this work has been funded by the European Research Council (ERC) under the European Union’s Horizon 2020 research and innovation programme (EAST project, grant agreement No. 864972).


\bibliographystyle{ACM-Reference-Format} 

%




\end{document}